\documentclass[aps,prl,twocolumn,showpacs,floatfix,superscriptaddress,longbibliography]{revtex4-1}
\usepackage{mathrsfs,longtable}
\usepackage{amssymb, amsbsy, amsmath, latexsym, dsfont, array, layout,mathrsfs,color,ulem,bm}
\usepackage[colorlinks,linkcolor=blue,anchorcolor=red,citecolor=blue]{hyperref}
\usepackage{multirow}
\usepackage{float}
\usepackage{threeparttable}
\usepackage{geometry}
\usepackage{ulem}
\usepackage{subfigure}
\newcommand{\one}{\mathds{1}}
\newcommand{\ket}[1]{\left|{#1}\right\rangle}
\newcommand{\bra}[1]{\left\langle{#1}\right|}

\usepackage{graphicx}
\usepackage{xspace, stmaryrd, ulem}
\definecolor{delete}{rgb}{1.0, 0.0, 0.0}
\definecolor{edit}{rgb}{0.0, 0.0, 0.9}
\definecolor{comment}{rgb}{0.9, 0.0, 0.0}

\DeclareMathOperator{\Tr}{Tr}

\begin{document}

\title{Experimental observation of the Yang-Lee quantum criticality in open systems}

\author{Huixia Gao}
\affiliation{Beijing Computational Science Research Center, Beijing 100084, China}
\author{Kunkun Wang}
\affiliation{School of Physics and Optoelectronic Engineering, Anhui University, Hefei 230601, China}
\author{Lei Xiao}
\affiliation{School of Physics, Southeast University, Nanjing 211189, China}
\affiliation{Beijing Computational Science Research Center, Beijing 100084, China}
\author{Masaya Nakagawa}
\author{Norifumi Matsumoto}
\affiliation{Department of Physics, University of Tokyo, 7-3-1 Hongo, Bunkyo-ku, Tokyo 113-0033, Japan}
\author{Dengke Qu}
\affiliation{Beijing Computational Science Research Center, Beijing 100084, China}
\author{Haiqing Lin}
\affiliation{School of Physics, Zhejiang University, Hangzhou 310030, China}
\author{Masahito Ueda}\email{ueda@cat.phys.s.u-tokyo.ac.jp}
\affiliation{Department of Physics, University of Tokyo, 7-3-1 Hongo, Bunkyo-ku, Tokyo 113-0033, Japan}
\affiliation{Institute for Physics of Intelligence, University of Tokyo, 7-3-1 Hongo, Bunkyo-ku, Tokyo 113-0033, Japan}
\affiliation{RIKEN Center for Emergent Matter Science (CEMS), Wako 351-0198, Japan}
\author{Peng Xue}\email{gnep.eux@gmail.com}
\affiliation{School of Physics, Southeast University, Nanjing 211189, China}
\affiliation{Beijing Computational Science Research Center, Beijing 100084, China}

\begin{abstract}
The Yang-Lee edge singularity was originally studied from the standpoint of mathematical foundations of phase transitions, and its physical demonstration has been of active interest both theoretically and experimentally. However, the presence of an imaginary magnetic field in the Yang-Lee edge singularity has made it challenging to develop a direct observation of the anomalous scaling with negative scaling dimension associated with this critical phenomenon.
We experimentally implement an imaginary magnetic field and demonstrate the Yang-Lee edge singularity through a nonunitary evolution governed by a non-Hermitian Hamiltonian in an open quantum system, where a classical system is mapped to a quantum system via the equivalent canonical partition function. In particular, we directly observe the partition function in our experiment using heralded single photons.  The nonunitary quantum criticality is identified with the singularity at an exceptional point.  We also demonstrate unconventional scaling laws for the finite-temperature dynamics unique to quantum systems.
\end{abstract}

\maketitle

{\it Introduction.---}Yang-Lee zeros~\cite{YL52,LY52} are the zero points of the partition function appearing only on the complex plane of the physical parameters and provide key properties of phase transitions, such as critical exponents~\cite{F78}. Yang and Lee~\cite{YL52,LY52} showed that zeros of the partition function of the classical ferromagnetic Ising model are distributed on the imaginary axis of the complex magnetic field~\cite{SG73,N74,LS81}. Yang-Lee zeros are also related to singularities~\cite{KG71,KF79,F80,C85,CM89,Z91,SV07,KLF21,VLF22,GKT17} in thermodynamic quantities. 
When the distribution of Yang-Lee zeros pinches (crosses) the real axis, the system exhibits a second-order (first-order) phase transition. Furthermore, the distribution itself exhibits a singularity at its edges, and such singularity is called the Yang-Lee edge singularity, 
which stands as a prototypical instance of nonunitary critical phenomena exhibiting  anomalous scaling laws unseen in unitary critical systems~\cite{CJS17,CYW20,IZ86,ISZ86}.

Due to their fundamental importance, Yang-Lee zeros and Yang-Lee edge singularity have been of  theoretical~\cite{DF19,DBF20,PBD21,H18,CJR20,JYB21} and experimental~\cite{JSH17,FVT18,BKK01,WL12,PZW15,BMP17,W17,W18,FZH21} interest. However, 
the presence of an imaginary magnetic field in the Yang-Lee edge singularity has made it challenging to develop a direct observation method and understand the physical implications of the anomalous scaling with negative scaling dimension associated with this phenomenon. 
Here the negative scaling dimension indicates that correlation functions diverge algebraically under space-time dilations and it is characteristic of nonunitary critical phenomena. A recent theoretical discovery demonstrates that the implementation of the Yang-Lee edge singularity in  quantum systems is achievable through the utilization of the quantum-classical correspondence~\cite{M76,K79}. This correspondence allows for the mapping of a classical system onto a quantum system by employing the equivalent canonical partition function ~\cite{MNU22}.

 In this Letter, we experimentally implement an imaginary magnetic field and demonstrate the Yang-Lee edge singularity  through a nonunitary evolution governed by a non-Hermitian Hamiltonian in an open quantum system.
  The Yang-Lee zeros and the Yang-Lee edge singularity of the classical ferromagnetic Ising model are both displayed in the quantum system due to the quantum-classical correspondence. The nonunitary quantum criticality is identified with the singularity at an exceptional point. We also show unconventional scaling laws for finite-temperature dynamics which are unique to quantum systems. Furthermore, we present the phase diagram of the Yang-Lee quantum critical system, where Yang-Lee zeros appear 
in the parity-time ($\mathcal{PT}$)-broken phase.  Our work is the first to measure all the critical
	exponents of the magnetization, magnetic susceptibility, two-time correlation function, and the density of zeros about the Yang-Lee edge singularity~\cite{sm}. In particular, we directly observe the partition function, which gives a crucial advantage in the study of Yang-Lee zeros and related topics.

{\it Yang-Lee edge singularity in open quantum systems.---}We consider the Yang-Lee edge singularity in the classical one-dimensional Ising model with a pure-imaginary magnetic field $H=-J\sum_j\sigma_j\sigma_{j+1}-ih\sum_j\sigma_j$~\cite{F80},
where $J>0$, $h\in \mathbb{R}$, and $\sigma_j=\pm1$.
This model can be mapped to a quantum system governed by a $\mathcal{PT}$-symmetric non-Hermitian Hamiltonian via the quantum-classical correspondence~\cite{BB98,BBF02}
\begin{align}
\label{eq:HPT}
	H_\mathcal{PT}=R\cos\phi\sigma_x+iR\sin\phi\sigma_z,
\end{align}
where $R>0$, $\phi \in (-\pi/2, \pi/2)$, and $\sigma_x$ and $\sigma_z$ are the Pauli matrices. 
The canonical partition function of $H$ is derived using the path-integral representation outlined in Eq.~(\ref{eq:HPT}) for its quantum counterpart. On the basis of 
the partition functions' equivalence, Matsumoto {\it et al.}~\cite{MNU22} pointed out that the latter quantum system exhibits a criticality equivalent to the Yang-Lee edge singularity in the former classical system. The Hamiltonian $H_\mathcal{PT}$ satisfies the $\mathcal{PT}$ symmetry with $\left[H,\mathcal{PT}\right]=0$, where $\mathcal{P}=\sigma_x$, $\mathcal{T}=\mathcal{K}$, and $\mathcal{K}$ is complex conjugation. The eigenenergies of the Hamiltonian are $E_{\pm} = \pm R \sqrt{\cos 2\phi}$. The exceptional points~\cite{B04,H12} are located at $\phi=\pm \pi/4$, separating the $\mathcal{PT}$-unbroken and  broken regimes.

The Yang-Lee edge singularity occurs at the edges of the distribution of zeros of the partition function
\begin{align}
	Z =
	\Tr\left[ e^{-\beta H_\mathcal{PT}} \right]
	= \sum_{p=\pm} e^{-\beta E_{p}},
\end{align}
where $\beta$ is the inverse temperature. The Yang-Lee quantum critical phenomena appear in the expectation value of a certain observable $O$ given by~\cite{UPJ79,G91,YHL17,ZHW20,MNU22}
\begin{align} \label{EV-PT}
	\hspace{-2mm}
	\left \langle O \right \rangle
	&= \frac{\Tr\left[ O e^{-\beta H_\mathcal{PT}} \right]}{Z}=\frac{1}{Z} \sum_{p=\pm} \frac{ \bra{E^{L}_{p}}{O}\ket{E^{R}_{p}}}{\left \langle E^{L}_{p}|E^{R}_{p} \right \rangle} e^{-\beta E_{p}},
\end{align}
where $\ket {E^{R}_{p}}$ ($\bra {E^{L}_{p}}$) is the right (left) eigenvector of $H_\mathcal{PT}$. We simulate the $\mathcal{PT}$-symmetric nonunitary quantum dynamics using a single-photon interferometric network, and experimentally investigate the Yang-Lee quantum criticality. 

\begin{figure}
	\includegraphics[width=0.5\textwidth]{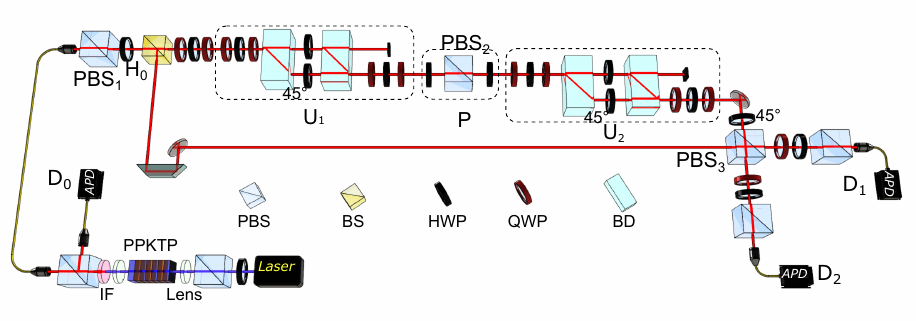}
	\caption{Experimental setup. Heralded single photons are created via  type-II spontaneous parametric down-conversion. The polarization beam splitter (PBS$_1$) and the half-wave plate (HWP) $H_0$ are used to generate initial polarization states $\ket{H}$ and $\ket{V}$. After photons pass through a $50:50$ beam splitter (BS), the transmitted photons enter the evolution path, and the reflected photons act as reference photons and interfere with the transmitted photons at PBS$_3$. The evolution process is divided into four parts. The elimination of the global phase is realized by wave plates. The nonunitary evolutions $U_1$ and $U_2$ are realized by two beam displacers (BDs) and sets of wave plates. The projector $P$ is realized by two HWPs and PBS$_2$. The measurement part is realized by PBS$_3$, two quarter-wave plates (QWPs) and two HWPs. Finally, photons are detected by avalanche photodiodes (APDs), and recording the coincidence counts of D$_0$, D$_1$ and D$_0$, D$_2$, respectively.}
	\label{fig:setup}
\end{figure}

{\it Experimental demonstration.---}To simulate the dynamics of the two-level $\mathcal{PT}$-symmetric system governed by $H_\mathcal{PT}$,  we employ as the basis states  the horizontal and vertical polarization states of a heralded single photon, i.e., $\{\ket{0}=\ket{H},\ket{1}=\ket{V}\}$. 
 Instead of implementing the non-Hermitian Hamiltonian $H_{PT}$, we simulate the nonunitary quantum dynamics by directly implementing a nonunitary  time-evolution operator $U$ such that $U=e^{-iH_\text{eff}t}$ at any given time $t$ (see Eq.~(\ref{nonunitary operator})). Here the effective non-Hermitian Hamiltonian is given by
\begin{align}
	H_\text{eff}=H_\mathcal{PT}+\frac{d}{t} \one,
\end{align}
where $d=i\ln{\frac{1}{\sqrt{\max|\lambda|}}}$, $\lambda$ is the eigenvalue of $e^{-iH_\mathcal{PT}t}e^{iH_\mathcal{PT}^\dagger t}$~\cite{H50,SLM18}, and $\one$ is a $2 \times 2$ identity matrix. The probability amplitudes with respect to $H_\text{eff}$ and $H_\mathcal{PT}$ are related to each other by
$\bra{j} e^{-i H_\text{eff}t}\ket{j}=\bra{j} e^{-i H_\mathcal{PT}t}\ket{j}/\sqrt{\max|\lambda|}$, where $j=H,V$.

As illustrated in Fig.~\ref{fig:setup}, the nonunitary operator $U$  is implemented on the basis of the following decomposition:
\begin{equation} \label{nonunitary operator}
U=R(\phi_2,\theta_2,\phi'_2)L(\theta_H,\theta_V)R(\phi_1,\theta_1,\phi'_1),
\end{equation}
where the rotation $R(\phi_j,\theta_j,\phi'_j)$ ($j=1,2$) can be realized by a set of sandwiched wave plates with a configuration quarter-wave plate (QWP) at $\phi_j$, a HWP at $\theta_j$ and a QWP at $\phi'_j$, and the polarization-dependent loss operator $L$ is realized by a combination of two beam displacers (BDs) and two HWPs with setting angles $\theta_H$ and $\theta_V$.  For each given evolution time $t$, the nonunitary evolution $U$ can be realized by tuning the setting angles of wave plates~\cite{XWZ+19} and  mapped to $e^{-iH_\mathcal{PT}t}$ with a correction factor $\sqrt{\max|\lambda|}$.

We characterize scaling laws of physical quantities for a finite-temperature quantum system via the magnetization
\begin{equation}
m=\langle \sigma_z\rangle=\frac{\bra{H} e^{-\beta H_\mathcal{PT}}\ket{H}-\bra{V} e^{-\beta H_\mathcal{PT}}\ket{V}}{\bra{H} e^{-\beta H_\mathcal{PT}}\ket{H}+\bra{V} e^{-\beta H_\mathcal{PT}}\ket{V}},
\end{equation}
the magnetic susceptibility
\begin{equation}
\chi=\frac{\text{d}m}{\text{d}a}=\frac{m-m'}{\tan\phi-\tan\phi'}
\end{equation} with $a=\tan\phi$ representing a normalized magnetic field and $m$ ($m'$) representing the magnetization for $H_\mathcal{PT}(\phi)$ ($H_\mathcal{PT}(\phi')$), and the two-time correlation function
\begin{align}\label{G}
&G(t_2,t_1)=\langle\sigma_z(t_2)\sigma_z(t_1)\rangle-\langle\sigma_z(t_2)\rangle\langle\sigma_z(t_1)\rangle\\
&=\frac{1}{Z}\Big(\Sigma_{HH}\Sigma'_{HH}-\Sigma_{HV}\Sigma'_{VH}-\Sigma_{VH}\Sigma'_{HV}+\Sigma_{VV}\Sigma'_{VV}\Big)\nonumber\\
&-m^2,\nonumber
\end{align}
where $\Sigma_{ij}=\bra{i} e^{-i\Delta t H_\mathcal{PT}}\ket{j}$, $\Sigma'_{ij}=\bra{i} e^{(i\Delta t-\beta) H_\mathcal{PT}}\ket{j}$ ($i,j=H,V$) and $Z=\bra{H} e^{-\beta H_\mathcal{PT}}\ket{H}+\bra{V} e^{-\beta H_\mathcal{PT}}\ket{V}$ is the partition function. Especially, the  finite-temperature scaling of $G(t_2,t_1)$ is unique to quantum critical phenomena.


 To measure the physical quantities experimentally, we use interference-based measurement. As illustrated in Fig.~\ref{fig:setup}, after single photons pass through a beam splitter (BS), the transmitted photons as signal photons go through a nonunitary evolution, while the reflected photons as reference photons remain unchanged and then interfere with the transmitted ones after the evolution at a PBS. Via QWPs, HWPs and PBSs, projective measurements with the bases of $\{\ket{+}, \ket{-}, \ket{R}\}$ ($\ket{\pm}=(\ket{H}\pm\ket{V})/\sqrt{2}$, $\ket{R}=(\ket{H}-i\ket{V})/\sqrt{2}$) are then performed on the polarizations of the photons transmitted and reflected by the PBS. 
The outputs are recorded in coincidence with trigger photons. Typical measurements yield a maximum of $240,000$ photon counts per second. For example, to measure $m$, we need to obtain both $\bra{H} e^{-\beta H_\mathcal{PT}}\ket{H}$ and $\bra{V} e^{-\beta H_\mathcal{PT}}\ket{V}$. First, photons are prepared in the initial state $\ket{H}$ (or $\ket{V}$). After signal photons undergo a nonunitary evolution via $U=e^{-iH_\text{eff}t}=e^{-\beta H_\mathcal{PT}}/\sqrt{\max|\lambda|}$ (here we take $t=-i\beta$), they interfere with the reference photons in $\ket{H}$ (or $\ket{V}$) at the PBS. The inverse temperature $\beta$ is taken as a parameter of the nonunitary evolution and tuned by the setting angles of wave plates. 
The overlap $\bra{H(V)} e^{-\beta H_\mathcal{PT}}\ket{H(V)}$ can be calculated by the coincidence counts~\cite{sm}. Similarly, we can obtain the  overlaps $\Sigma_{ij}\Sigma'_{ji}$ in $G(t_2,t_1)$ by applying the nonunitary operations $U_1=e^{-i\Delta t H_\text{eff}}= e^{-i\Delta t H_\mathcal{PT}}/\sqrt{\max|\lambda|}$ and $U_2=e^{(i\Delta t-\beta) H_\text{eff}}=e^{(i\Delta t-\beta)H_\mathcal{PT}}/\sqrt{\max|\lambda|}$ and the projector $P=\ket{j}\bra{j}$ on the signal photons.

\begin{figure}
	\includegraphics[width=0.5\textwidth]{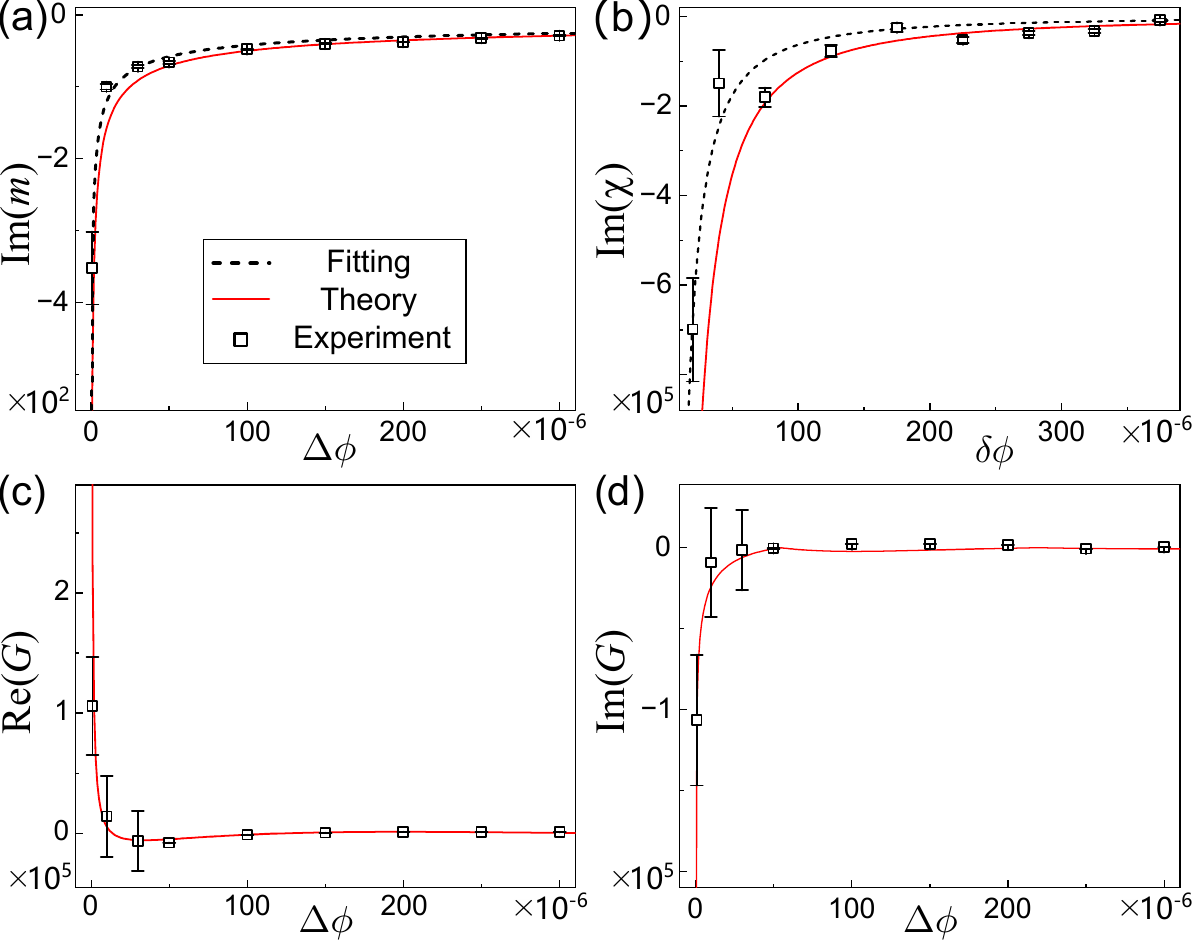}
	\caption{Yang-Lee scaling laws of physical quantities for a finite-temperature quantum system in the $\mathcal{PT}$-unbroken phase in the limit $\phi\to \pi/4-0$ after $\beta^{-1}\to0$. (a) The imaginary parts of $m$ as a function of $\Delta \phi$. (b) The imaginary parts of $\chi$ as a function of $\delta\phi=(\Delta\phi+\Delta\phi')/2$, where $\Delta\phi=(10,30,50,100,\cdots,350)\times10^{-6}$ and $\Delta\phi'=(30,50,100,150,\cdots,400)\times10^{-6}$. Dependencies of real (c) and imaginary (d) parts of $G(t_{2}, t_{1})$ on $\Delta \phi$. Experimental data are shown as open squares and theoretical predictions are represented by solid curves. We choose $R=0.05$, $\beta=10^5$ and $\Delta t=3000$. Dashed curves in (a) and (b) correspond to the results fitted by different power laws. Error bars indicate the statistical uncertainty, which are obtained from Monte Carlo simulations under the assumption of Poissonian photon-counting statistics. Some error bars are smaller than the size of the symbols.
	}
	\label{fig:data1}
\end{figure}

\begin{figure}[t]
	\includegraphics[width=0.5\textwidth]{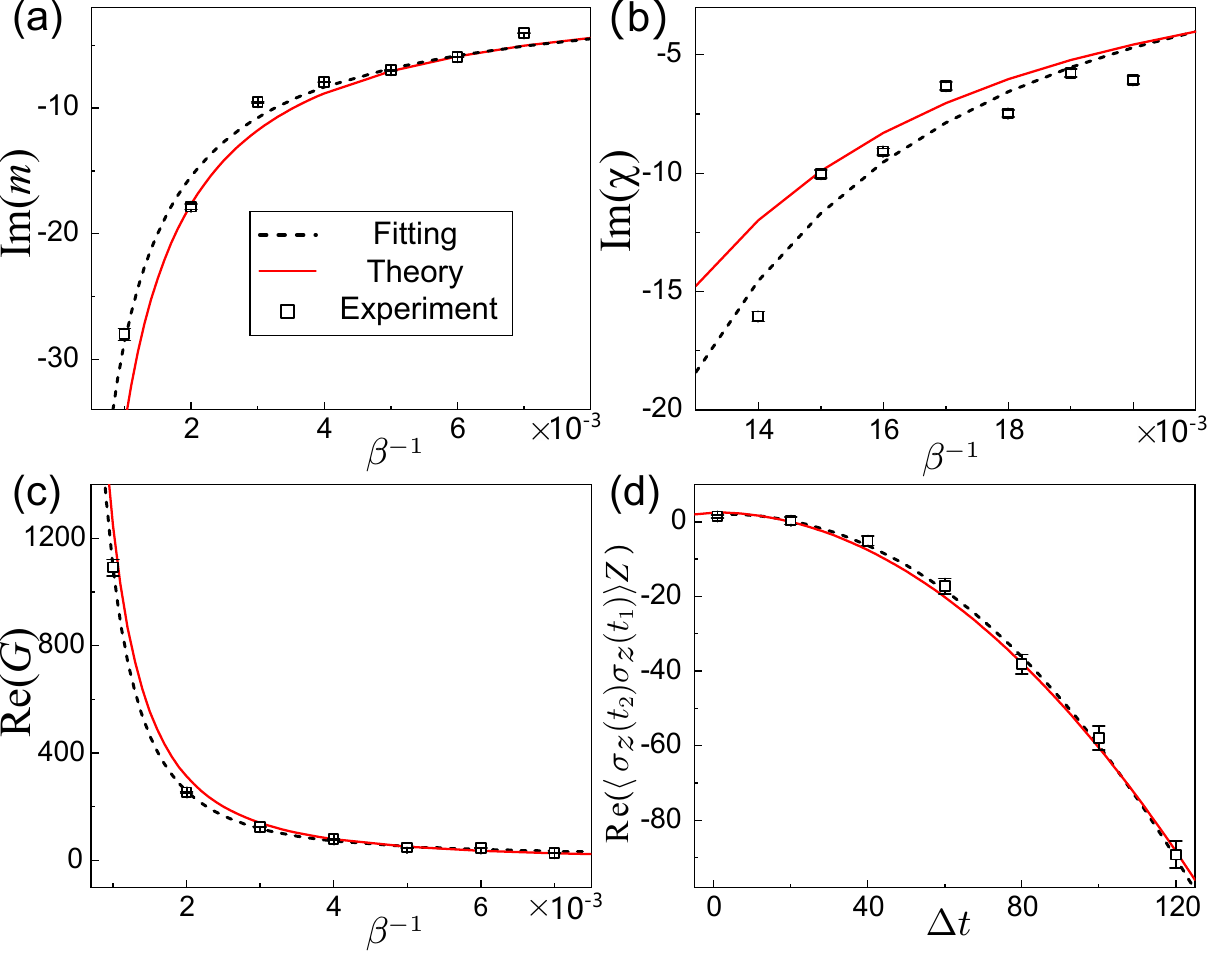}
	\caption{ Temperature dependences of  anomalous scaling laws and $\Delta t$ dependence of $\text{Re}(\langle\sigma_z(t_2)\sigma_z(t_1)\rangle Z)$ in the limit of $\beta^{-1}\to0$ after $\phi\to \pi/4-0$. The imaginary parts of $m$ (a) and $\chi$ (b) as functions of $\beta^{-1}$. (c) Real part of $G(t_{2}, t_{1})$ as a function of $\beta^{-1}$. (d) Real part of $\langle\sigma_z(t_2)\sigma_z(t_1)\rangle Z$ as a function of $\Delta t$. We choose $R=0.05$, $\phi=\pi/4-10^{-6}$ ($\phi'=\pi/4-10^{-2}$), $\Delta t =0.1$ in (a), (b) and (c), and  $R=0.05$, $\phi=\pi/4-10^{-6}$, $\beta=10^{4}$ in (d). 
}
	\label{fig:data2}
\end{figure}


\begin{figure}
	\includegraphics[width=0.5\textwidth]{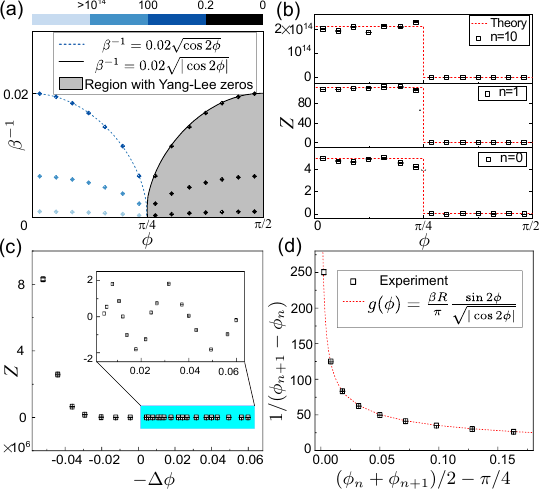}
	\caption{ (a) Phase diagram of the Yang-Lee quantum critical system. Experimental values of  $Z$ on the $\beta^{-1}-\phi$ plane. The critical point is located at $\phi=\pi/4$ and $\beta^{-1}=0$. 
The Yang-Lee zeros appear in the $\mathcal{PT}$-broken ($\phi>\pi/4$) phase (the grey region)  given by Eq.~(\ref{eq:range}
). In the $\mathcal{PT}$-unbroken phase, a crossover between Yang-Lee scaling laws and unconventional scaling laws occurs around the region indicated by the blue dashed curve.   
The experimental data are obtained with $n=0$, $n=1$, $n=10$ 
from top to bottom, and the color indicates the value of $Z$. (b) $Z$ as a function of $\phi$ with $n=10$, $n=1$ and $n=0$
, respectively. 
We choose $R=0.01\pi$ for (a-b). For (c-d) we choose $R=0.05$ and $\beta=10^3$. (c) $Z$ versus $-\Delta\phi$.  The region colored in light blue shows the $\mathcal{PT}$-broken regime, and the inset is an enlarged view of the regime.
(d) Density of zeros $1/(\phi_{n+1}-\phi_{n})$  for  $n=0,1,2,\cdots,9$. The horizontal axis is taken as $(\phi_{n}+\phi_{n+1})/2$, which is measured from the critical point $\phi=\pi/4$. 
}
	\label{fig:data4}
\end{figure}

{\it Yang-Lee critical phenomena.---}First, we discuss the scaling laws of the system in the $\mathcal{PT}$-unbroken phase ($|\phi|<\pi/4$) by examining the dependence of the physical quantities on the parameter $\Delta\phi:=\pi/4-\phi$. We consider two cases, in which we take either  the limit $\phi\rightarrow \pi/4-0$ after $1/\beta\rightarrow 0$ or the limit  $1/\beta\rightarrow 0$ after $\phi\rightarrow \pi/4-0$.  Here $\beta^{-1}= 0$  corresponds to the thermodynamic limit of the classical one-dimensional Ising model, and the exceptional points $\phi = \pm \pi/4$ separate the $\mathcal{PT}$-unbroken and $\mathcal{PT}$-broken phases.

For the former case, the scaling laws are equivalent to those in the classical  counterpart~\cite{MNU22,F80}, i.e.,
\begin{align}
&m\rightarrow \frac{-i\sin\phi}{\sqrt{\cos2\phi}}\propto\Delta\phi^{-\frac{1}{2}}, \text{ }\text{ }\text{ }\text{ } \chi\rightarrow\frac{-i\cos^{3}\phi}{\cos^{\frac{3}{2}}2\phi}\propto \Delta\phi^{- \frac{3}{2}}, \nonumber\\
&G(t_2,t_1)\rightarrow\frac{\cos^{2}\phi}{\cos2\phi}\exp\left[- 2\pi i \frac{\Delta t}{\pi/(R\sqrt{\cos2\phi})}\right].
\end{align}
 We  experimentally test these scaling laws for $R=0.05$, $\beta=10^{5}$, and various $\Delta \phi$. Since the real parts of $m$ and $\chi$ are zeros, the scaling laws of them with respect to $\Delta \phi$ are characterized by their imaginary parts in Figs.~\ref{fig:data1}(a) and (b), respectively. 
By fitting the power exponents to $m \sim \Delta \phi^{r}$, $\chi \sim \delta \phi^{r'}$,  we obtain $r\sim-0.458\pm0.033$, $r'\sim-1.477\pm0.350$, which agree with their theoretical predictions $-0.5$ and $-1.5$, respectively.  The discrepancy with the theoretical predictions is mainly because the fitting result is sensitive to the  difference between the leftmost experimental data point in Fig.~\ref{fig:data1}(a) and its theoretical prediction,  where the slope of the curve becomes large. 
In Figs.~\ref{fig:data1}(c) and (d), we also show the measured  values of $G(t_2,t_1)$ with respect to $\Delta \phi$ for $\Delta t=3000$,  which agree with the theoretical predictions~\cite{sm}. 

For the latter case, 
we study the scaling laws
\begin{align}
\label{anomalous}
&m \to  - \frac{i}{\sqrt{2}} \beta R, \text{ }\text{ }\text{ }\text{ }\chi\to  - \frac{i} {3\sqrt{2}}  (\beta^{3} R^{3} + \frac{3}{2} \beta R ),
\nonumber\\
&G( t_{2}, t_{1} )
	\to R^{2}\left[\frac{1}{2}\beta^{2} -  i \beta \Delta t - (\Delta t)^{2}\right] + 1,
\end{align}
which have not been discussed in the classical systems~\cite{F80,MNU22}. The critical exponents for the power-law dependence on the temperature $\beta^{-1}$ are $-1$, $-3$ and $-2$, respectively. In addition, the two-time correlation function scales as $G(t_{2}, t_{1})\propto(\Delta t)^{2}$ in the limit of $\Delta t \to \infty$. If $\Delta t$ is replaced by an imaginary-time interval $-i \Delta \beta$, $G(t_2,t_1)$ is equivalent to the spatial correlation function $G_{cl}(x)$ of the classical system with the distance  $x=\Delta \beta$. 
The power-law scaling $G_{\rm cl}(x) \propto x^{-2 \Delta}$ with a negative scaling dimension $\Delta = -1$ is consistent with the critical scaling in the corresponding classical system~\cite{F80}. To test this unconventional scaling laws, 
 we choose $R=0.05$, $\phi=\pi/4-10^{-6}$ ($\phi'=\pi/4-10^{-2}$), and  $\Delta t =0.1$.  By fitting the power exponents with the formula $m\sim (\beta^{-1})^{r}$, $\chi\sim a(\beta^{-1})^{r'}+b(\beta^{-1})^{r}$, we obtain $r\sim-0.890\pm0.011$ and  $r'\sim-3.171\pm0.062$, which agree with their theoretical predictions $-1$ and $-3$, respectively.   
As illustrated in Fig.~\ref{fig:data2}(c), by fitting the data with $G(t_2,t_1)\sim (\beta^{-1}) ^{r''}$, we obtain the experimental result $r''\sim-2.160\pm0.021$, which is consistent with the theoretical critical exponent $-2$.

 We also show the $\Delta t$ dependence of $\langle\sigma_z(t_2)\sigma_z(t_1)\rangle Z$ with $\beta=10^{4}$ in Fig.~\ref{fig:data2}(d),  which is equivalent to the dependence of $G(t_{2}, t_{1})$ because  $Z$ and $m$ are independent of $\Delta t$ [see Eq.~(\ref{G})].
The power exponent is fitted by the formula $\langle\sigma_z(t_2)\sigma_z(t_1)\rangle Z \sim (\Delta t)^{s}$, and the obtained fitting result is  $s\sim 2.187\pm0.089$,
which is consistent with the theoretical prediction $G(t_2,t_1) \sim (\Delta t)^{2}$.

{\it  Phase diagram and partition function.---}In the $\mathcal{PT}$-broken  phase ($|\phi| > \pi/4$),
$m$, $\chi$ and $G(t_2,t_1)$ diverge periodically in the limit $\phi \to \pi/4 + 0$ after $\beta^{-1}\to 0$~\cite{MNU22}. The corresponding experimental data can be found in the Supplemental Material ~\cite{sm}. The condition for the divergence is
\begin{equation} \label{scaling-broken}
	\beta R \sqrt{|\cos 2\phi|} =(n + \frac{1}{2}) \pi,
\end{equation}
where $n$ is an integer, which corresponds to the condition for zeros of the partition function
\begin{equation}\label{Yang-Lee zero}
	Z =  2 \cos(\beta R \sqrt{|\cos 2\phi|}),
\end{equation}
i.e., the Yang-Lee zeros. These zeros appear only in the region defined by
\begin{equation}
\label{eq:range}
\beta^{-1} \le \frac{2} {\pi} R \sqrt{|\cos 2\phi|},
\end{equation}
in the $\mathcal{PT}$-broken phase.



We measure the partition function $Z$ on the $\beta^{-1}-\phi$ plane with $R=0.01\pi$, $n=0$, $n=1$ and $n=10$ in Eq.~(\ref{scaling-broken}) for $14$ different $\phi$. As illustrated in Fig.~\ref{fig:data4}(a), the Yang-Lee zeros  appear only in the region given by Eq.~(\ref{eq:range}) in the $\mathcal{PT}$-broken phase, which agrees with the theoretical prediction.  As shown in Fig.~\ref{fig:data4}(b), $Z$ takes large positive values in the $\mathcal{PT}$-unbroken phase and drops in the $\mathcal{PT}$-broken phase. As illustrated in Fig.~\ref{fig:data4}(c), we can see $Z$ oscillating with $-\Delta\phi$ in the $\mathcal{PT}$-broken phase.  Since the nodes of the oscillations correspond to Yang-Lee zeros, the period of the oscillation reflects the distance between Yang-Lee zeros. We can utilize Eq.~(\ref{scaling-broken}) to understand the relationship between the period of the oscillation and the parameters $\beta$ and $R$. For $R = 0.05$ and $\beta = 10^3$, we expect the oscillation period of $\phi$ to be $\pi/100$, which is consistent with the observed value of $0.03$.
Thus, we observe the Yang-Lee zeros experimentally and also demonstrate that the Yang-Lee edge singularity indeed manifests itself as the distribution of zeros of $Z$.

{\it Density of zeros.---}For the $(0+1)$-dimensional quantum Yang-Lee model in Eq.~(\ref{eq:HPT}), the zero points $\{\phi_n\}_n$ of $Z$ are determined from the condition in Eq.~(\ref{scaling-broken}). The distribution of zeros of $Z$ becomes dense if the limit $\beta R\rightarrow \infty$ is taken. We then calculate the density of zeros $g(\phi):=\sum_n \delta(\phi-\phi_n)$ and find  a power-law behaviour
\begin{equation}
g(\phi)\propto (-\Delta\phi)^{-\frac{1}{2}}
\end{equation}
near the critical point $\phi=\pi/4$.  This result is consistent with the classical Yang-Lee edge singularity~\cite{F78,F80}. In Fig.~\ref{fig:data4}(d), we show the experimental results of the density of zeros $1/(\phi_{n+1}-\phi_{n})$  for $n = 0, 1, 2. \cdots, 9$, with the parameters $R=0.05$ and $\beta=10^3$. The horizontal axis is taken as $(\phi_{n}+\phi_{n+1})/2$, which is measured from the critical point $\phi=\pi/4$. Our experimental results agree well with the analytic expression of $g(\phi)=\frac{\beta R}{\pi}\frac{\sin2\phi}{\sqrt{|\cos2\phi|}}$~\cite{sm}. 
Via the density distribution of zeros experimentally observed for $10$ different $n$'s, we have achieved the direct observation of the Yang-Lee edge singularity.

{\it Conclusion.---} The Yang-Lee edge singularity is a quintessential nonunitary critical phenomenon characterized by anomalous scaling. In this Letter, we have experimentally demonstrated the Yang-Lee singularity in a non-Hermitian quantum system with $\mathcal{PT}$ symmetry. Specifically, we have observed both anomalous scaling laws that are consistent with the classical Yang-Lee singularity and unconventional scaling laws that have not been discussed in classical systems. In particular, we directly observe the partition function in our experiment, which gives a decisive advantage in the study of Yang-Lee zeros and related topics. Our work presents the first experimental demonstration of the Yang-Lee quantum criticality in open quantum systems. We expect that the nonunitary critical phenomena in open quantum systems for higher-dimensional systems can also be probed using a similar approach.

\subsection{ Appendix A: Density of zeros in the $(0+1)$-dimensional quantum Yang-Lee model}

We consider the $(0 + 1)$-dimensional quantum Yang-Lee model~\cite{MNU22} \begin{align}
\label{eq:HPT2}
H_\mathcal{PT}=R\cos\phi\sigma_x+iR\sin\phi\sigma_z,
\end{align}
where $R>0$, $\phi \in (-\pi/2, \pi/2)$, and $\sigma_x$ and $\sigma_z$ are the Pauli matrices. The partition function of this system in the $\mathcal{PT}$-broken regime ($|\phi|>\pi/4$) is given by $Z = \Tr[ e^{-\beta H_\mathcal{PT}} ]= 2 \cos(\beta R \sqrt{|\cos 2\phi|})$. The zero points $\{\phi_n\}_n$ of the partition function are determined from the condition
\begin{align}
\label{scaling-broken}
	\beta R \sqrt{|\cos 2\phi_n|} =(n + \frac{1}{2}) \pi,
\end{align}
where $n = 0, 1, 2,\cdots$.

The distribution of zeros of the partition function becomes dense if we take the limit $\beta R\rightarrow \infty$. The Yang-Lee edge singularity manifests itself as a power-law behavior of the density of zeros near the edge of the distribution~\cite{KG71,F78}. Here we calculate the density of zeros
\begin{equation}
	g(\phi):=\sum_n \delta(\phi-\phi_n)
\end{equation}
near the critical point $\phi=\pi/4$. To this end, we assume $\beta R\gg1$ and introduce a continuous variable $x$ that satisfies
\begin{align} \label{variable x}
	x:=\frac{\pi}{\beta R}n.
\end{align}
From Eq.~(\ref{scaling-broken}), the zero points $\phi(x)$ satisfy
\begin{equation}
	\cos 2\phi(x)=-(x+\frac{\pi}{2\beta R})^2
\end{equation}
and thus we have
\begin{equation}
	\frac{dx}{d\phi}=\frac{\sin 2\phi}{x+\pi/(2\beta R)}=\frac{\sin2\phi}{\sqrt{|\cos 2\phi|}}.
\end{equation}
Then, the density of zeros can be calculated as
\begin{align}\label{g}
	g(\phi^{'})&:=\sum_n \delta(\phi^{'}-\phi_n)\simeq\frac{\beta R}{\pi}\int dx\delta(\phi^{'}-\phi(x))\\ &=\frac{\beta R}{\pi}\int d\phi\frac{dx}{d\phi}\delta(\phi^{'}-\phi)=\frac{\beta R}{\pi}\frac{\sin2\phi^{'}}{\sqrt{|\cos 2\phi^{'}|}}.\nonumber
\end{align}
Thus, the density of zeros shows a power-law behavior near the critical point as
\begin{equation}
g(\phi)\propto (-\Delta\phi)^{\sigma},
\end{equation}
where $\Delta\phi:=\pi/4-\phi$ and $\sigma=-1/2$. The critical exponent $\sigma=-1/2$ agrees with the result for the classical one-dimensional Ising model ~\cite{F78,F80}.

\subsection{Appendix B: Measurement of the density of zeros in the Yang-Lee quantum criticality}

To observe the power-law behaviour in the density of zeros, we choose the parameters $R=0.05$ and $\beta=10^3$ and show the experimental results of the density of zeros $1/(\phi_{n+1}-\phi_{n})$ from $n=0$ to $9$ versus $(\phi_{n}+\phi_{n+1})/2-\pi/4$. First, we choose several $\phi$ and measure the partition function $Z$ experimentally. By fitting with $Z\sim \cos(\beta R\sqrt{|\cos2\phi|})$, we obtain the experimental result $\beta R \sim 49.682 \pm0.029$, which is consistent with the theoretical prediction of $50$. Second, for $n=0,\cdots,9$, we calculate $\phi_n$ from  Eq.~(11) of the main text and measure the corresponding partition function $Z_n$. With the measured $Z_n$ and $\beta R$, we obtain the experimental results of $\phi_n$ from
\begin{equation}
\phi_n=\frac{1}{2}\arccos\left[-\left(\frac{1}{\beta R}\arccos\left(\frac{Z_n}{2}\right)\right)^2\right].
\end{equation}
The experimental results of the density of zeros $1/(\phi_{n+1}-\phi_{n})$ from $n=0$ to $9$ are shown in  Fig.~4(d) of the main text. The horizontal axis is taken as $(\phi_{n}+\phi_{n+1})/2$, which is measured from the critical point $\phi=\pi/4$. Our experimental results agree well with the analytic expression  Eq.~(\ref{g}).
The Yang-Lee edge singularity manifests itself as the distribution of zeros for $10$ different $n$'s.

\subsection{Appendix C: Experimental implementation}

Here we provide the method for measuring the physical quantities including the magnetization, the magnetic susceptibility, the two-time correlation function and the partition function. As demonstrated in the main text, these physical quantities can be obtained by measuring the probabilistic amplitudes of the final states.

For example, if we want to obtain the term $\bra{0} e^{-\beta H_\mathcal{PT}} \ket{0}$ of the magnetization in  Eq.~(6) of the main text, we measure the probability amplitude $p_1$ which is related to the overlap between the initial state $\ket{0}$ and the final state after the nonunitary evolution $e^{-\beta H_\mathcal{PT}}$ as $\bra{0} e^{-\beta H_\mathcal{PT}} \ket{0}=p_1 \times \sqrt{\max|\lambda|}$ , where $\lambda$ is the eigenvalue of $e^{-\beta H_\mathcal{PT}}e^{-\beta H_\mathcal{PT}^\dagger}$. Once we obtain the probability amplitude $p_1$, we can calculate the measured overlap and the measured physical quantities.

\begin{figure}[h]
	\includegraphics[width=0.5\textwidth]{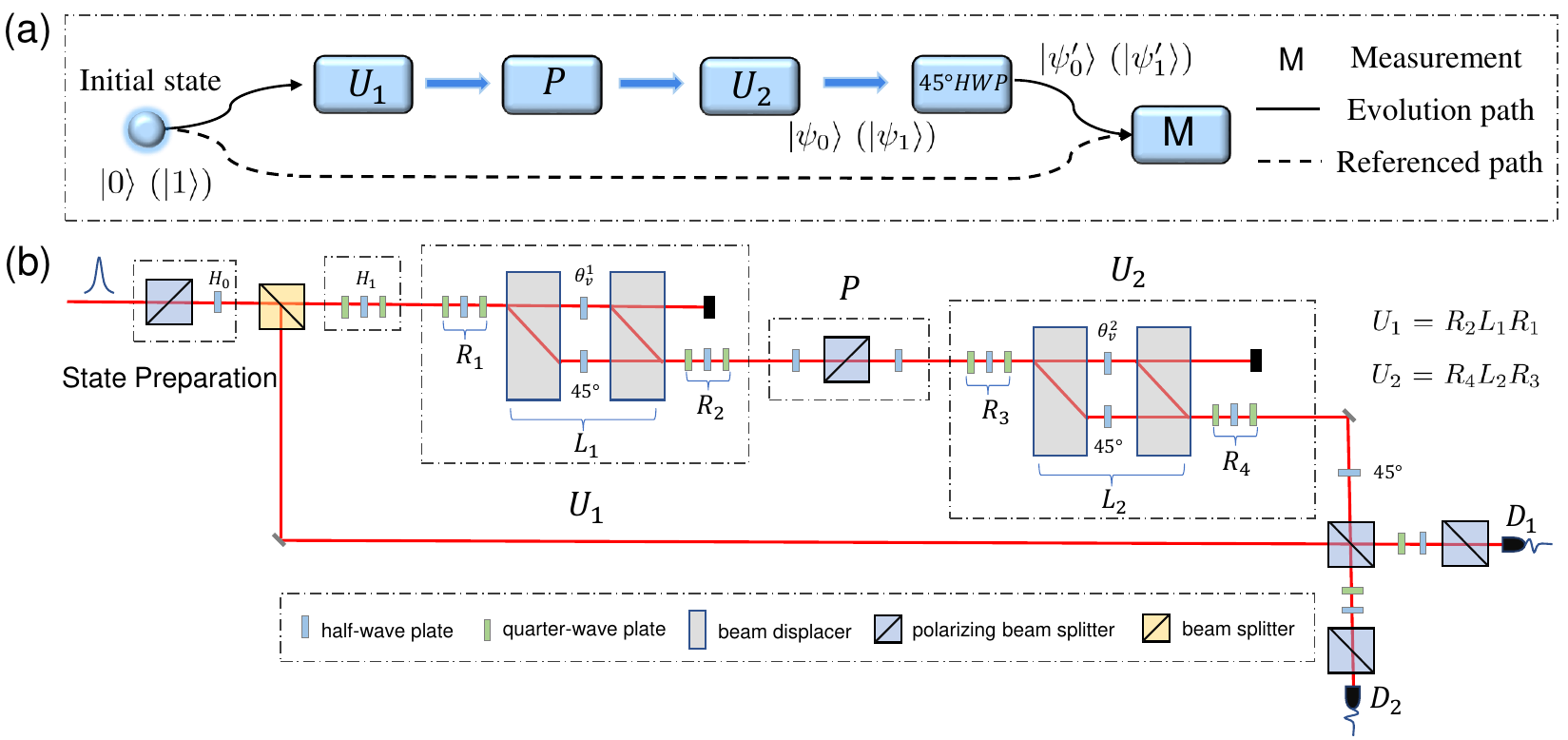}
	\caption{Schematic diagrams of the experiment. (a) The flow chart of state evolution. Single photons with the initial state $\ket{0}$ or $\ket{1}$ goes through two paths of state evolution and reference. On the evolution path, single photons evolves into $\ket{\psi_0}$ or $\ket{\psi_1}$ after undergoing non-unitary evolutions $U_1$, $P$, and $U_2$, and then evolves into $\ket{\psi'_0}$ or $\ket{\psi'_1}$ by passing through a half-wave plate (HWP) at $45^{\circ}$. The photons remain unchanged on the reference path. Finally, we make measurements on the photons after they interfere. (b) Experimental setup which is the same as Fig.~1 in the main text.
	}
	\label{fig:scheme}
\end{figure}

\begin{figure*}
	\includegraphics[width=\textwidth]{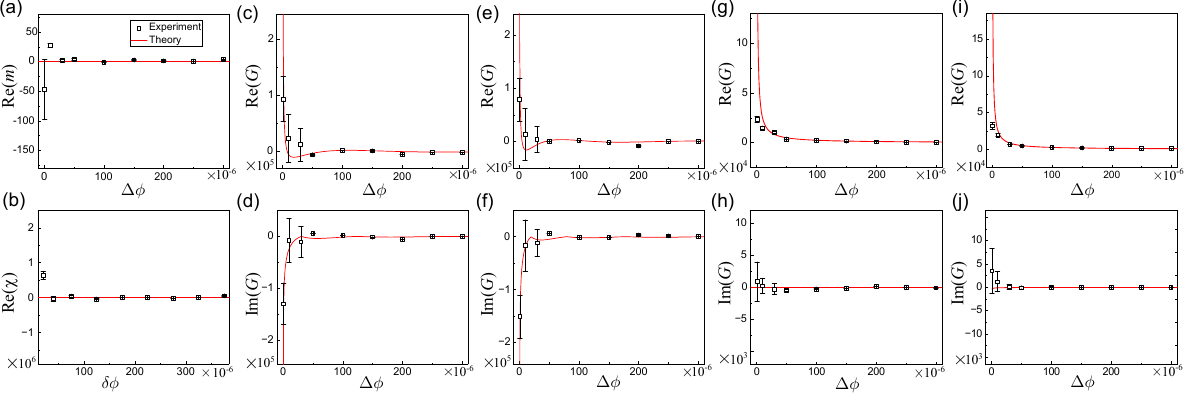}
	\caption{(a) Real parts of the expectation value of the magnetization  plotted against $\Delta \phi$. (b) Real parts of the expectation value of the magnetic susceptibility as a function of $\delta\phi=(\Delta\phi+\Delta \phi')/2$, where $\Delta\phi=10\times10^{-6},30\times10^{-6},50\times10^{-6},100\times10^{-6},\cdots,350\times10^{-6}$ and $\Delta\phi'=30\times10^{-6},50\times10^{-6},100\times10^{-6},150\times10^{-6},\cdots,400\times10^{-6}$. Real (c) and imaginary (d) parts of expectation values of the two-time correlation function for different $\Delta \phi$ with $\Delta t =4000$. Real (e) and imaginary (f) parts of expectation values of the two-time correlation function plotted against $\Delta \phi$ with $\Delta t =5000$.   Real (g) and imaginary (h) parts of expectation values of the two-time correlation function plotted against $\Delta \phi$ with $\Delta t =0.1$.  Real (i) and imaginary (j) parts of expectation values of the two-time correlation function plotted against $\Delta \phi$ with $\Delta t =10$. We choose $R=0.05$ and $\beta=10^{5}$. Black open squares indicate experimental data and red solid curves show theoretical predictions.
	}
	\label{fig:dataS1}
\end{figure*}

\begin{figure}
	\includegraphics[width=0.5\textwidth]{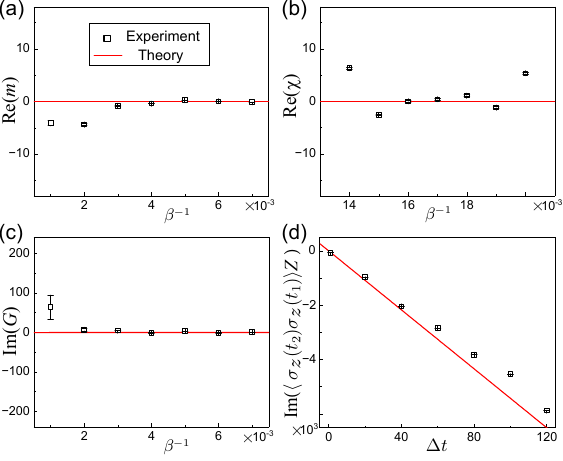}
	\caption{ (a) Real parts of the expectation value of the magnetization  plotted against $\beta^{-1}$. (b) Real parts of the expectation value of the magnetic susceptibility  plotted against $\beta^{-1}$. (c) Imaginary parts of expectation values of the two-time correlation function  plotted against $\beta^{-1}$. (d) Imaginary parts of expectation values of $\langle\sigma_z(t_2)\sigma_z(t_1)\rangle Z$  plotted against $\Delta t$.  We choose $R=0.05$, $\phi=\pi/4-10^{-6}$ ($\phi'=\pi/4-10^{-2}$), $\Delta t =0.1$ in (a), (b) and (c), and  $R=0.05$,  $\phi=\pi/4-10^{-6}$, $\beta=10^{4}$ in (d). Black open squares indicate experimental data and red solid curves show theoretical predictions.}
	
	\label{fig:dataS2}
\end{figure}

\begin{figure}[t]
	\includegraphics[width=0.35\textwidth]{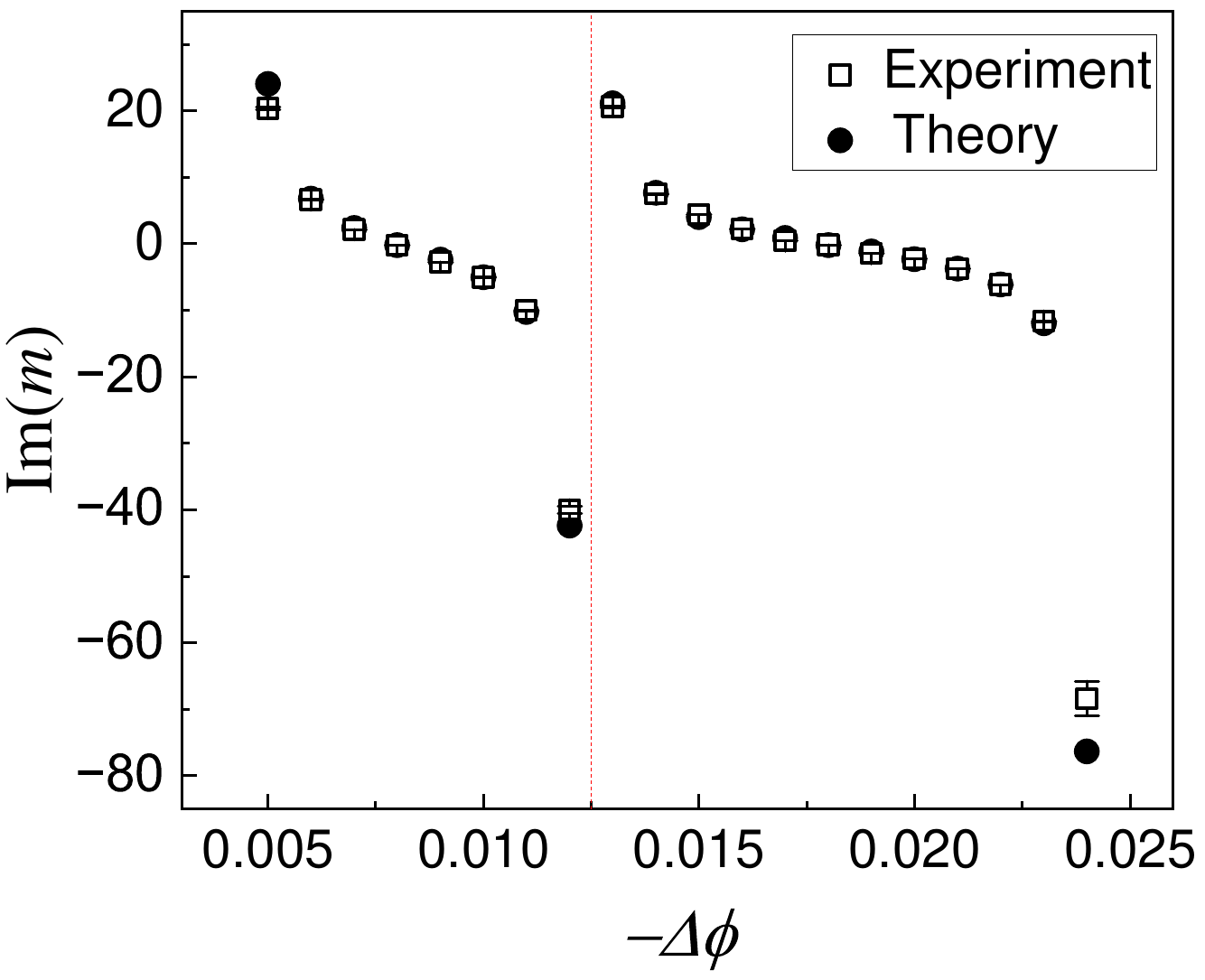}
	\caption{Imaginary parts of the magnetization against $-\Delta \phi=\phi-\pi/4$ in the $\mathcal{PT}$-broken phase. They are plotted over two periods and separated by the red dashed line. We choose $R=0.05$ and $\beta=10^{3}$. Error bars indicate the statistical uncertainty, which are obtained from Monte Carlo simulations under the assumption of Poissonian photon-counting statistics. Some error bars are smaller than the size of the symbols.
	}
	\label{fig:data3}
\end{figure}

\begin{figure}
	\includegraphics[width=0.4\textwidth]{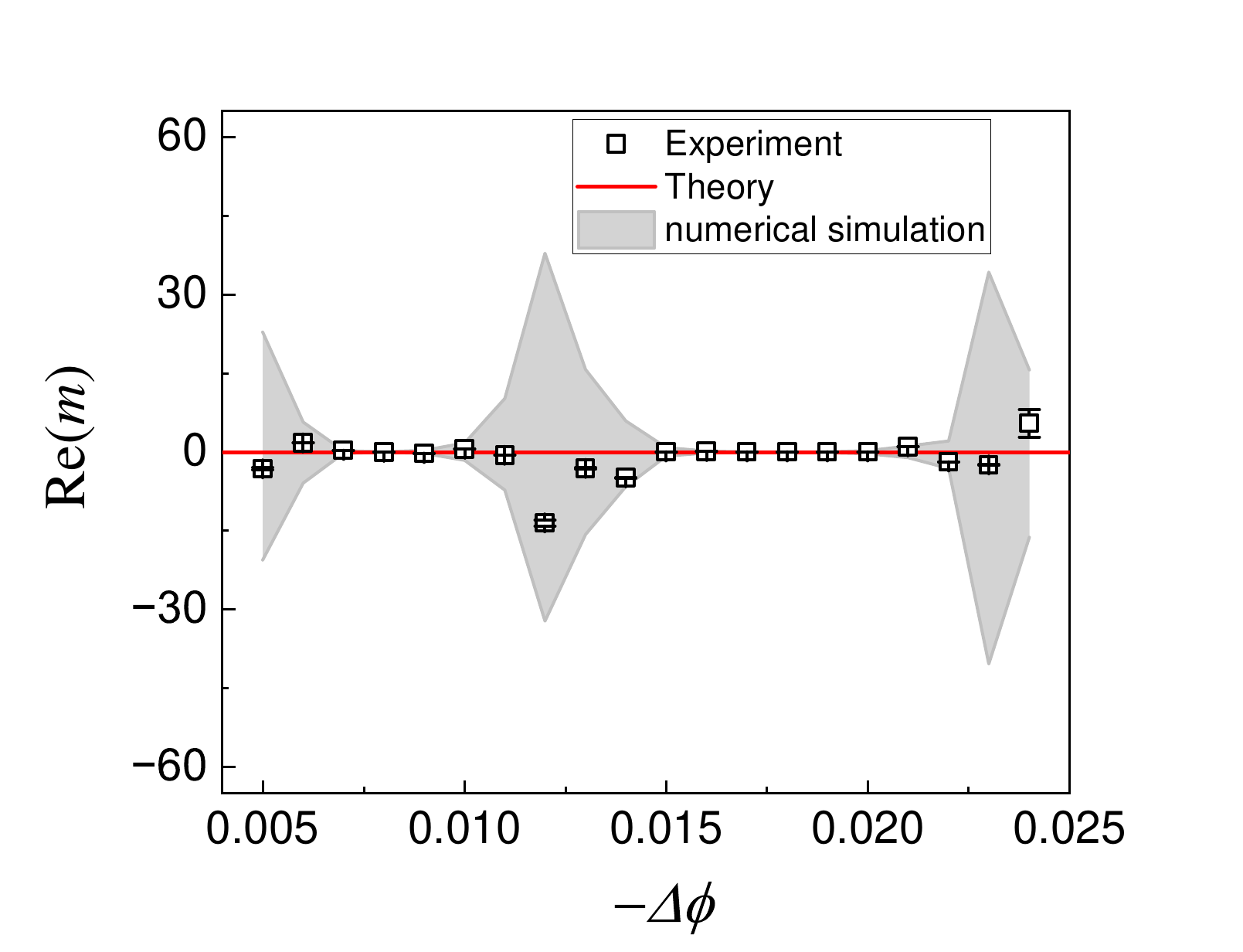}
	\caption{Real part of the expectation value of the magnetization plotted against $-\Delta \phi$. We choose $R=0.05$, $\beta=10^3$. Open squares indicate experimental data, the solid line shows the theoretical prediction. Error bars indicate the statistical uncertainty, which are obtained from Monte Carlo simulations under the assumption of Poissonian photon-counting statistics. Some error bars are smaller than the size of the symbols. Shadows indicate systematic errors which are obtained from the numerical simulation by considering the inaccuracy of wave plates and decoherence in the experiment.
	}
	\label{fig:dataS3}
\end{figure}

Figure~\ref{fig:scheme} shows the flow chart of state evolution and the schematic diagram of state evolving elements. In our experiment, a polarizing beam splitter (PBS) and a half-wave plate (HWP) H$_0$ realize the preparation of the initial state $\ket{0}$ or $\ket{1}$. After the heralded single photons pass through the $50:50$ beam splitter (BS), the transmitted photons go through the nonunitary evolution and the reflected photons  remain unchanged. The state of the transmitted photons then evolves according to $\ket{\psi_0}=e^{-\beta H_\mathcal{PT}} \ket{0}=\begin{pmatrix} a_0 \\ b_0 \end{pmatrix}$ or $\ket{\psi_1} =e^{-\beta H_\mathcal{PT}} \ket{1}= \begin{pmatrix} a_1 \\ b_1 \end{pmatrix}$. In the nonunitary evolution process, the elimination of the global phase is made by a HWP H$_1$ sandwiched between two quarter-wave plates (QWPs at $45^{\circ}$). To realize nonunitary evolution $U_1$, we perform the singular value decomposition on the nonunitary matrix $U_1=R_2 L_1 R_1$~\cite{xwz19}.  The rotations $R_1$ and $R_2$ can be realized  by a sandwich-type set of wave plates including a HWP and two QWPs. The loss operator $L_1=\begin{pmatrix}
	0 & \sin2\theta_{v}^{1}\\
	1 & 0
\end{pmatrix}$
is implemented by two HWPs at $45^{\circ}$ and $\theta^1_v$, respectively, and two beam displacers (BDs). Similarly, we can realize another nonunitary evolution $U_2$ with the same method. For the projector $P$,  $\ket{0} \bra{0}$ can be realized by a PBS, while $\ket{1} \bra{1}$ can be realized by a PBS and two HWPs at $45^\circ$. A HWP at $45^{\circ}$ is applied to the polarizations of the transmitted photons. Then the state evolves into $\ket{\psi'_0}=\begin{pmatrix}b_0 \\ a_0 \end{pmatrix}$ or $\ket{\psi'_1} = \begin{pmatrix} b_1 \\ a_1 \end{pmatrix}$. The reflected photons act as reference photons and interfere with transmitted photons at the third PBS.

We perform projective measurements~\cite{XDW19,WXB21} in the basis $\left\{\ket{+}, \ket{-}, \ket{R}\right\}$ on the polarizations of the photons in either of the transmitted ($t$) or reflected mode ($r$) of the third PBS. We denote the coincidence counts of D$_0$ and D$_1$ (D$_0$ and D$_2$) as $N^{r}$ ($N^{t}$) when the initial state is $\ket{0}$ ($\ket{1}$), and  the measured coincidence counts are denoted as $\left\{N^{r}_+, N^{r}_-, N^{r}_R\right\}$ ($\left\{N^{t}_+, N^{t}_-, N^{t}_R\right\}$), which satisfy the following relations
 \begin{align}
 	N^{r}_+ &= \mathcal{N}\frac{a^{2}_0+1+a^{*}_0+a_0}{2}, \quad
 	N^{r}_- = \mathcal{N}\frac{a^{2}_0+1-a^{*}_0-a_0}{2},  \nonumber \\
 	N^{r}_R &= \mathcal{N}\frac{a^{2}_0+1+i(a_0-a^{*}_0)}{2},\quad
 	N^{t}_+ = \mathcal{N}\frac{b^{2}_1+1+b^{*}_1+b_1}{2},  \nonumber \\
 	N^{t}_- &= \mathcal{N}\frac{b^{2}_1+1-b^{*}_1-b_1}{2}, \quad
 	N^{t}_R = \mathcal{N}\frac{b^{2}_1+1+i(b^{*}_1-b_1)}{2},
\end{align}
where $\mathcal{N}$ is the number of reference photons. We can obtain the probability amplitudes $a_0$ and $b_0$ through the following relations:
\begin{align}
	&{\rm Re} (a_0)=\frac{N^{r}_+ -N^{r}_- }{2\mathcal{N}}, \quad
	{\rm Im}(a_0)=\frac{1}{\mathcal{N}}\times \frac{N^{r}_+ +	N^{r}_- -2N^{r}_R}{2},\nonumber \\
	&{\rm Re}(b_1)=\frac{N^{t}_+ -N^{t}_- }{2\mathcal{N}}, \quad
	{\rm Im}(b_1)=\frac{1}{\mathcal{N}}\times \frac{2	N^{t}_R - (N^{t}_+ +N^{t}_-) }{2}.
\end{align}
Then, we can calculate the measured overlap and the measured physical quantities  from the real and imaginary parts of the amplitudes of the evolved states.

\subsection{Appendix D: Additional experimental results}

In this section, we present the additional part of the complex expectation values of observables presented in the main text.


The first task of our experiment in the main text is to find the imaginary parts of  expectation values of the magnetization and the magnetic susceptibility for different values of $\phi$ [see Figs.~2(a) and (b) in the main text]. As shown in Figs.~\ref{fig:dataS1}(a) and (b), the real parts of their expectation values are zero. In addition, we show the real and imaginary parts of the two-time correlation function for $\Delta \phi=\pi/4-\phi$ with $\Delta t=4000$, $\Delta t=5000$,  $\Delta t=0.1$ and $\Delta t=10$ in Figs.~\ref{fig:dataS1}(c), (d), (e), (f), (g), (h), and (i), (j), respectively. It can be seen that the measured two-time correlation functions are consistent with their theoretical predictions for different time differences $\Delta t$.

The second task of our experiment in the main text is to find the imaginary parts of expectation values of the magnetization and the magnetic susceptibility for different values of $\beta^{-1}$ [see Figs.~3(a) and (b) in the main text]. As shown in Figs.~\ref{fig:dataS2}(a) and (b),  the real parts of their  expectation values are zero. In addition, for the expectation value of the two-time correlation function, we plot the imaginary parts against $\beta^{-1}$ and $\Delta t$ in Figs.~\ref{fig:dataS2}(c) and (d), respectively. It can been seen that the measured two-time correlation functions are consistent with their theoretical predictions.

	Here we comment on the choice of the parameters. In this work, we investigate two types of scaling laws, i.e., the Yang-Lee scaling laws [Eq. (9) in the main text] and the anomalous scaling laws [Eq. (10) in the main text]. Since the two scaling behaviors are separated by a crossover line $\beta^{-1}=(2R / \pi)\sqrt{\cos 2\phi}$ (see the phase diagram in Fig. 4(a) in the main text and Fig. 1 in Ref. \cite{MNU22}), the former should be observed if $\beta^{-1} \ll (2R / \pi)\sqrt{\cos 2\phi}$ and $\phi < \pi/4$, while the latter should be observed if $(2R / \pi)\sqrt{|\cos 2\phi|} < \beta^{-1} \ll 2R / \pi$. The parameters in Figs. 2 and 3 are chosen so that these conditions are satisfied. In particular, in Fig. 3, we are concerned with the anomalous scaling behavior which becomes prominent at low temperatures. Therefore, if we make the value of $\pi/4-\phi$ fixed to be a smaller one to examine the regime of smaller values of $\beta^{-1}$, we expect to obtain the critical exponents closer to the theoretical predictions. Furthermore, the $\beta^{-1}$ dependence of the two-time correlation function in Fig. 3(c) is expected to exhibit values of critical exponents closer to the theoretical predictions in the regime $\beta\gg\Delta t$.


In the $\mathcal{PT}$-broken phase, we experimentally confirm that the expectation value of the magnetization is pure imaginary for several different values of $\Delta \phi$. The third task shows imaginary parts of the expectation value of the magnetization in the $\mathcal{PT}$-broken phase in Fig.~\ref{fig:data3}. We examine the dependence of the magnetization on $\phi$ by taking the limit $\phi \to \pi/4 + 0$ after $\beta^{-1} \to 0$. We examine the magnetization as an example for $R=0.05$, $\beta=10^{3}$, and $20$ different values of $\phi$. Since the real parts of the magnetization is zero, we only show its imaginary parts in two variation periods in Fig.~\ref{fig:data3}. The magnetization exhibits a discontinuity with respect to $-\Delta\phi$. The experimental results demonstrate that there is no scaling laws for the magnetization in the $\mathcal{PT}$-broken phase.

We show in Fig.~\ref{fig:dataS3} the real part of the expectation value of the magnetization for several different values of $-\Delta \phi$. The measured data are consistent with the theoretical predictions except for a few points that deviate from zero presumably due to systematic errors.

\begin{table}
	\caption{The squared correction factor $\max|\lambda|$. The squared correction factor $\max|\lambda_Z|$ needs to be multiplied when measuring the partition function $Z$. The squared correction factor $\max|\lambda_G|$ needs to be multiplied when measuring $\langle\sigma_z(t_2)\sigma_z(t_1)\rangle Z$ to obtain the two-time correlation function $G(t_2,t_1)$.
	}
	\renewcommand{\arraystretch}{1.3}
	\setlength{\tabcolsep}{1mm}
	
	\subtable[The squared correction factors $\max|\lambda_Z|$ and $\max|\lambda_G|$ in Fig. 2 of the main text and Fig.~\ref{fig:dataS1}. The subscripts $1,2,3,4,5$ denote the different choices  $\Delta t=0.1$, $\Delta t=10$, $\Delta t=3000$, $\Delta t=4000$ and $\Delta t=5000$, respectively.
	]{\scalebox{0.42}{
		\begin{tabular}{cccccccccccc}
			\hline
			\bfseries  $\Delta \phi/ 10^{-6}$   &  $1$&  $10$&  $30$ &  $50$ & $100$ & $150$ & $200$ &  $250$ & $300$    \\
			\hline
			
			\bfseries $\max|\lambda_Z|$ &$3.466\times 10^{11}$&$6.610\times10^{23}$&$3.640\times10^{37}$& $1.344\times 10^{47}$ & $6.555\times 10^{64}$& $2.780\times 10^{78}$& $9.036\times 10^{89}$& $1.292\times 10^{100}$ & $2.000\times 10^{109}$  \\
			\hline
			
			\bfseries $(\max|\lambda_G|)_1$ &$3.490\times 10^{11}$ &$6.657\times10^{23}$ &$3.666\times10^{37}$& $1.354\times 10^{47}$ & $6.601\times 10^{64}$& $2.799\times 10^{78}$& $9.101\times 10^{89}$& $1.302\times 10^{100}$ & $2.014\times 10^{109}$ \\
			\hline
			
			\bfseries $(\max|\lambda_G|)_2$ &$6.931\times10^{11}$ &$1.322\times10^{24}$ &$7.281\times10^{37}$& $2.688\times 10^{47}$ & $1.311\times 10^{65}$& $5.559\times 10^{78}$& $1.807\times 10^{90}$& $2.584\times 10^{100}$ & $3.999\times 10^{109}$ \\
			\hline
			
			\bfseries $(\max|\lambda_G|)_3$ &$1.536\times 10^{16}$&$2.554\times 10^{28}$&$1.022\times 10^{42}$& $2.675\times 10^{51}$ & $4.761\times 10^{68}$& $4.960\times 10^{81}$& $9.174\times 10^{91}$& $2.325\times 10^{102}$ & $1.722\times 10^{112}$ \\
			\hline
			
			\bfseries $(\max|\lambda_G|)_4$ &$2.699\times 10^{16}$&$4.020\times10^{28}$&$1.213\times10^{42}$& $2.223\times 10^{51}$ & $6.233\times 10^{67}$& $1.867\times 10^{81}$& $2.588\times 10^{93}$& $4.877\times 10^{103}$ & $6.438\times 10^{112}$ \\
			\hline
			\bfseries $(\max|\lambda_G|)_5$ &$4.155\times 10^{16}$&$5.345\times10^{28}$&$1.058\times10^{42}$& $9.631\times 10^{50}$ & $9.668\times 10^{67}$& $1.595\times 10^{82}$& $4.155\times 10^{93}$& $2.112\times 10^{103}$ & $1.720\times 10^{111}$ \\
			\hline
		\end{tabular}
		\label{firsttable}
	}}
	
	\qquad
	
	\subtable[The squared correction factors $\max|\lambda_Z|$ and $\max|\lambda_G|$ in Fig.~3 of the main text.]{\scalebox{0.5}{
		\begin{tabular}{ccccccccccc}
			\hline
			\bfseries  $\beta^{-1}\times10^{-3}$   &  $1$ &  $2$ & $3$ & $4$ & $5$ &  $6$ & $7$    \\
			\hline
			
			\bfseries $\max|\lambda_Z|$ &$5010.350$& $1252.520$ & $557.658$& $314.530$& $202.009$& $140.888$ & $104.035$  \\
			\hline
			
			\bfseries $\max|\lambda_G|$ &$5045.900$& $1261.410$ & $561.615$& $316.762$& $203.442$& $141.888$ & $104.773$\\
			\hline
			\hline
			\bfseries  $\Delta t$   &  $1$ &  $20$ & $40$ & $60$ & $80$ &  $100$ & $120$    \\
			\hline
			
			\bfseries $\max|\lambda_Z|$ &$5.000\times10^{5}$& $5.000\times10^{5}$ & $5.000\times10^{5}$& $5.000\times10^{5}$& $5.000\times10^{5}$& $5.000\times10^{5}$ & $5.000\times10^{5}$  \\
			\hline
			\bfseries $\max|\lambda_G|$ &$5.366\times10^{5}$& $1.866\times10^{6}$ & $4.950\times10^{6}$& $9.975\times10^{6}$& $1.699\times10^{7}$& $2.599\times10^{7}$ & $3.700\times10^{7}$\\
			\hline
		\end{tabular}
		\label{secondtable}
	}}

	\subtable[The squared correction factor $\max|\lambda_Z|$ in  Figs. 4(a) and (b) of the main text.]{\scalebox{0.5}{
		\begin{tabular}{cccccccccccc}
			\hline
			\bfseries  $\phi/(\pi/28)$ ($n=0$)  &  $0.5$ &  $1.5$ & $2.5$ & $3.5$ & $4.5$ &  $5.5$ & $6.5$    \\
			\hline
			
			\bfseries $\max|\lambda_Z|$ &$23.208$& $23.772$ & $25.061$& $27.535$& $32.470$& $44.639$ & $107.183$  \\
			\hline
			\hline
			\bfseries  $\phi/(\pi/28)$ ($n=1$)  &  $0.5$ &  $1.5$ & $2.5$ & $3.5$ & $4.5$ &  $5.5$ & $6.5$   \\
			\hline
			
			\bfseries $\max|\lambda_Z|$ &$12430.800$& $12759.900$ & $13513.000$& $1457.600$& $17840.500$& $24953.000$ & $61525.500$   \\
			\hline
			\hline
			\bfseries  $\phi/(\pi/28)$ ($n=10$)  &  $0.5$ &  $1.5$ & $2.5$ & $3.5$ & $4.5$ &  $5.5$ & $6.5$   \\
			\hline
			
			\bfseries $\max|\lambda_Z|$ &$4.502\times10^{28}$& $4.620\times10^{28}$ & $4.894\times10^{28}$& $5.416\times10^{28}$& $6.460\times10^{28}$& $9.034\times10^{28}$ & $2.228\times10^{29}$  \\
			\hline
			\hline
			\bfseries  $\phi/(\pi/28)$  &  $7.5$ &  $8.5$ & $9.5$ & $10.5$ & $11.5$ &  $12.5$ & $13.5$   \\
			\hline
			
			\bfseries $\max|\lambda_Z|$ &$17.807$& $5.886$ & $3.471$& $2.414$& $1.809$& $1.409$ & $1.119$  \\
			\hline
		\end{tabular}
		\label{thirdtable}
	}}
	
	\subtable[The squared correction factor $\max|\lambda_Z|$ in  Figs. 4(c) and (d) of the main text.]{\scalebox{0.45}{
		\begin{tabular}{cccccccccccc}
			\hline
			\bfseries  $-\Delta\phi $   &  $-0.052$ &  $-0.044$ & $-0.036$ & $-0.029$ & $-0.020$ & $-0.012$ & $-0.004$ &  $0.004$
			\\
			\hline
			\bfseries $\max|\lambda_Z|$ &$5.228\times10^{14}$&$	4.638\times10^{13}$&$	3.430\times10^{12}$&$	2.814\times10^{11}$& $7.196\times10^{9}$&	$1.347\times10^{8}$&$	6.785\times10^{5}$ &$227.077$ \\
			\hline
			\hline
			\bfseries  $-\Delta\phi $    &  $0.006$ & $0.008$ & $0.011$ & $0.012$ & $0.014$ & $0.018$ &  $0.022$ &  $0.024$
			\\
			\hline
			\bfseries $\max|\lambda_Z|$  &$135.424$ &$	1.026$ &$69.321$ & $81.274$ &	$54.709$ &$1.066$ &$34.384$ &$41.311$ \\
			\hline
			\hline
			\bfseries  $-\Delta\phi $    & $0.027$ & $0.032$ & $0.037$ & $0.040$ & $0.043$&  $0.049$ &  $0.056$ & $0.060$
			\\
			\hline
			\bfseries $\max|\lambda_Z|$  &$28.328$ &$1.002$ & $20.693$ &	$24.985$ &$17.674$&$1.018$ &$13.850$ &$16.673$ \\
			\hline
			\hline
			\bfseries  $n $   &  $0$ &  $1$ & $2$ & $3$ & $4$ & $5$ & $6$ & $7$ & $8$ & $9$
			\\
			\hline
			\bfseries $\max|\lambda_Z|$ &$2026.420$ &$225.154$ &$81.045$ &$41.331$ & $24.978$ &	$16.687$ &$11.907$  &$8.894$ &$6.866$ &$5.429$ \\
			\hline
		\end{tabular}
		\label{fourthtable}
	}}
	
	\label{tb:s}

\end{table}

\subsection{Appendix E: Error analysis}

Two factors are responsible for the deviation between the experimental results and their theoretical predictions. On one hand, the imperfection of the experiment affects the results of the experiment. On the other hand, the value of the correction factor $\sqrt{\max |\lambda|}$ amplifies the imperfection of the experiment.

There are three major sources of imperfections in our experiment: photon fluctuations, inaccuracy of wave plates and decoherence (dephasing) caused by the imperfect interferometers. First, the imperfection caused by photon-number fluctuations decreases with increasing photon counts. In our experiment, the coincidence count per second is about $240,000$, and the coincidence window is $150$s. It can be seen from the data plots that the error bars caused by photon number fluctuations in our experiments are small. Second, for each wave plate, we assume an uncertainty in the setting angle $\theta+\delta\theta$. Here $\delta\theta$ is randomly chosen from the interval $\left[-1^{\circ},1^{\circ}\right]$ (normal accuracy range of wave plates). Third, the dephasing due to the imperfection of interferometric network with PBSs (BDs) affects the experimental results through a noisy channel $\varepsilon(\rho)=\eta\rho+(1-\eta)\sigma_z\rho\sigma_z$ ($\varepsilon(\rho)=\eta_1\rho+(1-\eta_1)\sigma_z\rho\sigma_z$) acting on the polarization state in different output modes. Here $\rho$ ($\varepsilon(\rho)$) is the density matrix of the input (output) state of the noisy channel. In our experiments, the dephasing rate $\eta$ ($\eta_1$) for the interferometric network with PBSs (BDs) is $\eta=0.9615$ ($\eta_1=0.9995$).
As illustrated in Fig.~\ref{fig:dataS3}, shadows indicate systematic errors which are obtained from the numerical simulation by considering the inaccuracy of wave plates and decoherence in the
experiment. We found that the deviation between the experimental data and their theoretical predictions can be explained by the systematic errors.

As we can only implement passive non-unitary operations $U$ with only loss, we need to map them to the desired ones $e^{-iH_\mathcal{PT}t}$ via a correction factor $\sqrt{\max|\lambda|}$. Thus, for the experimental results of some physical quantities, we also need to correct them by multiplying the correction factor $\sqrt{\mathrm{max}|\lambda|}$.

As shown in Fig.~\ref{fig:dataS3}, the difference between the experimental data of the magnetization and their theoretical predictions is caused by the imperfection of the experiment. The difference between the experimental data of the two-time correlation function and the partition function and their theoretical predictions is not only related to the imperfection of the experiment, but also related to the squared correction factor $\max|\lambda|$. With the same experimental imperfection, a larger $\max|\lambda|$ magnifies the difference between the experimental data and their theoretical prediction. We show the squared correction factor $\max|\lambda|$ for each data in Table~\ref{tb:s}.

\subsection{Appendix F: Correction factors}

The correction factor $\lambda$ is crucial for the precise mapping between $H_\text{eff}$ and $H_\mathcal{PT}$. There are two correction parameters $\lambda_{Z}$ and $\lambda_{G}$. The first one $\lambda_{Z}$ is the eigenvalue of $e^{-\beta H_\mathcal{PT}}e^{-\beta H_\mathcal{PT}^\dagger}$. The second one satisfies $\lambda_{G}=\lambda_{G_1}\lambda_{G_2}$, where $\lambda_{G_1}$ is the eigenvalue of $e^{-i\Delta t H_\mathcal{PT}} e^{i\Delta t H_\mathcal{PT}^\dagger}$, and $\lambda_{G_2}$ is the eigenvalue of $e^{(i\Delta t-\beta) H_\mathcal{PT}} e^{(-i\Delta t-\beta) H_\mathcal{PT}^\dagger}$. The partition function $Z$ and the measured $Z_\text{eff}$ are related to each other by $Z=\sqrt{\max|\lambda_Z|} Z_\text{eff}$. Similarly, the correction factor $\sqrt{\max|\lambda_G|}$ needs to be multiplied when measuring $\langle\sigma_z(t_2)\sigma_z(t_1)\rangle Z$ to obtain the two-time correlation function $G(t_2,t_1)$. Whereas, neither of the magnetization and the magnetic susceptibility needs to be corrected by the correction factors.


The effect of the correction factors on the measured values of the magnetization, the magnetic susceptibility, the partition function and the two-time correlation function are shown  as
\begin{align}
&m=\langle \sigma_z\rangle=\frac{\bra{0} e^{-\beta H_\mathcal{PT}}\ket{0}-\bra{1} e^{-\beta H_\mathcal{PT}}\ket{1}}{\bra{0} e^{-\beta H_\mathcal{PT}}\ket{0}+\bra{1} e^{-\beta H_\mathcal{PT}}\ket{1}} \nonumber \\
&=\frac{\sqrt{\max|\lambda_Z|}\bra{0} e^{-\beta H_\text{eff}}\ket{0} -\sqrt{\max|\lambda_Z|}\bra{1} e^{-\beta H_\text{eff}}\ket{1} }{\sqrt{\max|\lambda_Z|}\bra{0} e^{-\beta H_\text{eff}}\ket{0} +\sqrt{\max|\lambda_Z|}\bra{1} e^{-\beta H_\text{eff}}\ket{1}}\nonumber\\
&=\frac{\bra{0} e^{-\beta H_\text{eff}}\ket{0} -\bra{1} e^{-\beta H_\text{eff}}\ket{1}}{\bra{0} e^{-\beta H_\text{eff}}\ket{0} +\bra{1} e^{-\beta H_\text{eff}}\ket{1}}\nonumber \\
&=m_\text{eff},
\end{align}
\begin{equation} \chi=\frac{\text{d}m}{\text{d}a}=\frac{m-m'}{\tan\phi-\tan\phi'}=\frac{m_\text{eff}-m'_\text{eff}}{\tan\phi-\tan\phi'}=\chi_\text{eff},
\end{equation}
\begin{align}
&Z=\bra{0} e^{-\beta H_\mathcal{PT}}\ket{0}+\bra{1} e^{-\beta H_\mathcal{PT}}\ket{1}\nonumber \\
&=\sqrt{\max|\lambda_Z|}\bra{0} e^{-\beta H_\text{eff}}\ket{0}+\sqrt{\max|\lambda_Z|}\bra{1} e^{-\beta H_\text{eff}} \ket{1} \nonumber\\
&=\sqrt{\max|\lambda_Z|}(\bra{0} e^{-\beta H_\text{eff}}\ket{0}+\bra{1} e^{-\beta H_\text{eff}}\ket{1})\nonumber \\
&=\sqrt{\max|\lambda_Z|} Z_\text{eff},
\end{align}
and
\begin{align}\label{G} G(t_2,t_1)=&\langle\sigma_z(t_2)\sigma_z(t_1)\rangle-\langle\sigma_z(t_2)\rangle\langle\sigma_z(t_1)\rangle\nonumber \\
	=&\frac{1}{Z}(\langle\sigma_z(t_2)\sigma_z(t_1)\rangle Z)-m^2\nonumber\\
	=&\frac{\sqrt{\max|\lambda_G|} (\langle\sigma_z(t_2)\sigma_z(t_1)\rangle Z)_\text{eff}}{\sqrt{\max|\lambda_Z|} Z_\text{eff}}-m_\text{eff}^2,
\end{align}
where
\begin{align}
	(\langle\sigma_z(t_2)\sigma_z(t_1)\rangle Z)_\text{eff}&=\bra{0} e^{-i\Delta t H_\text{eff}}\ket{0}\bra{0} e^{(i\Delta t-\beta) H_\text{eff}}\ket{0}\nonumber\\
	&-\bra{0} e^{-i\Delta t H_\text{eff}}\ket{1}\bra{1} e^{(i\Delta t -\beta) H_\text{eff}}\ket{0}\nonumber\\
	&-\bra{1} e^{-i\Delta t H_\text{eff}}\ket{0}\bra{0} e^{(i\Delta t-\beta) H_\text{eff}}\ket{1}\nonumber\\
	&+\bra{1} e^{-i\Delta t H_\text{eff}}\ket{1}\bra{1} e^{(i\Delta t-\beta) H_\text{eff}}\ket{1}.
\end{align}
Thus, the correction factor $\lambda$ only affects the partition function and two-time correlation function, but has no effect on the magnetization and the magnetic susceptibility. We note that the huge correction factor in Tables~\ref{tb:s}(a) and (b) is mostly cancelled in forming the ratio between $\sqrt{\max|\lambda_G|}$ and $\sqrt{\max|\lambda_Z|}$ in Eq.~(\ref{G}).

\subsection{Appendix G: Comparison with previous works}

We compare the experimental results of the Yang-Lee zeros and edge singularity phenomena in this work with other previous experimental works as illustrated in Table~\ref{comparison_table}. We emphasize that our work is the first to measure all the critical exponents of the magnetization, magnetic susceptibility, two-time correlation function, and the density of zeros about the Yang-Lee edge singularity. Measurements of the entire set of critical exponents provide the key to the determination of the universality class of the underlying physics.  Specifically, Ref. \cite{BKK01} estimates the dependence on the imaginary magnetic field via mathematical analytic continuation from measurements results for the real magnetic field. This reference measured the critical exponent of the density distribution of Yang-Lee zeros but did not measure other critical exponents. References \cite{ PZW15, BMP17, FVT18, FZH21} only measured the locations of Yang-Lee zeros but did not measure critical exponents.
 	
In contrast, our work directly controls the pure imaginary magnetic field via photon loss, and has the distinct advantages in the sense that we can directly measure the partition function, Yang-Lee zeros, and the physical quantities under the imaginary magnetic field.  In particular, two-time correlation functions can only be measured by our method. This is a consequence of the fact that our method has the decisive ability to perform real-time simulations under the imaginary magnetic field by using open systems.
 	
We also stress that the temperature dependence of the Yang-Lee critical phenomenon can only be measured by our method. In this sense, our work experimentally realizes the Yang-Lee quantum criticality beyond the classical regime.

\begin{table}[h]
	\renewcommand{\arraystretch}{1.3}
	\caption{ Comparison of Yang-Lee zeros and edge singularity phenomena observed in this work with other previous experimental works. }
	
	\label{comparison_table}
	\centering
	
	\resizebox{0.5\textwidth}{!}{
		\setlength{\tabcolsep}{0.5mm}{
			\begin{tabular}{m{1cm}|c|c|c|c|c|c|}
				\hline
				\multicolumn{1}{|c|}{\multirow{2}{*}{\bfseries Ref.}} &\multirow{2}{*}{\bfseries Yang-Lee zeros}  &\multicolumn{4}{c|} {\bfseries Critical exponents}
				\\
				\cline{3-6}
				
				\multicolumn{1}{|c|}{}& &magnetization&susceptibility&correlation functions&the density distribution of zeros \\
				\hline
				\multicolumn{1}{|c|}{\cite{BKK01} } &yes&no&no&no&yes\\
				\hline
				\multicolumn{1}{|c|}{\cite{PZW15} } &yes&no&no&no&no\\
				\hline
				\multicolumn{1}{|c|}{\cite{BMP17} } &yes&no&no&no&no\\
				\hline
				\multicolumn{1}{|c|}{\cite{FVT18} } &yes&no&no&no&no\\
				\hline
				\multicolumn{1}{|c|}{\cite{FZH21} } &yes&no&no&no&no\\
				\hline			
				\multicolumn{1}{|c|}{our work} &yes&yes&yes&yes&yes\\
				\hline

	\end{tabular}}}
	\label{tb1}
\end{table}

\begin{acknowledgments}
This work is supported by the National Natural Science Foundation of China (Grant Nos.~92265209 and 12025401). MU acknowledges support by KAKENHI Grant No. JP22H01152 from the Japan Society for the Promotion of Science  and by the CREST program ``Quantum Frontiers" (Grant No. JPMJCR23I1) by the Japan Science and Technology Agency. MN is supported by KAKENHI Grant No. JP20K14383 from the Japan Society for the Promotion of Science. NM is supported by KAKENHI Grant No. JP21J11280 from the Japan Society for the Promotion of Science. HQL acknowledges support from the National Natural Science Foundation of China (Grant No.~12088101).
\end{acknowledgments}

\end{document}